\documentclass[useAMS,usenatbib]{mn2e} 
\usepackage{graphicx,epsfig} 
\usepackage{epsf}
\usepackage{aas_macros}
\usepackage{appendix}
\usepackage[usenames]{color}

\begin{document}

\title[Some characteristics of \textit{Kepler} Short and Long Cadence Data]{An examination of some characteristics of \textit{Kepler} Short and Long Cadence Data}

\author[Simon J. Murphy] 
{Simon J. Murphy\\ 
\\
Jeremiah Horrocks Institute, University of Central Lancashire, Preston PR1 2HE, UK; smurphy6@uclan.ac.uk}

\maketitle

\begin{abstract}
A close comparison of \textit{Kepler} short- and long-cadence data released prior to 2011\,Nov\,1 has shown some subtle differences that make the short-cadence data superior to their long-cadence counterparts. The inevitable results of a faster sampling rate are present: the short-cadence data provide greater time resolution for short-lived events like flares, and have a much higher Nyquist frequency than the long-cadence data; however, they also contain fewer high-amplitude peaks at low frequency and allow a more precise determination of pulsation frequencies, amplitudes and phases. The latter observation indicates that \textit{Kepler} data are not normally distributed. Moreover, a close inspection of the Pre-search Data Conditioned (PDC) long-cadence data show residuals that have increased noise on time-scales important to asteroseismology, but unimportant to planet searches.
\end{abstract}


\section{Introduction}

The Kepler Space Telescope is in a 372.5-d heliocentric Earth-trailing orbit, collecting white-light photometric data for a sample of $\sim$160\,000 stars covering a field of view of 115 deg$^{2}$. The core goal of the mission is detection of Earth-like planets orbiting Sun-like stars within the habitable zone. The details of the mission goals and design are described by \citet{kochetal2010} and \citet{boruckietal2010}.

Secondary to the main goal of the mission is asteroseismology, which can provide valuable information on the host stars that is important in planet characterisation. With transit depths, only the ratio of the planet and host star radii is available, but asteroseismology allows detailed analysis of the star's interior if the star pulsates, and for solar-like oscillators can yield the star's radius to better than 3\,per\,cent \citep{stelloetal2009} -- even approaching 1\,per\,cent in some cases \citep{gillilandetal2010a}. Through asteroseismology, \textit{Kepler} also promises significant advances in stellar astrophysics with the dedication of $\sim$1\,per\,cent of observations to asteroseismic study \citep{gillilandetal2010a}.

\textit{Kepler} data are available in two cadences, long (LC) and short (SC). Each cadence is composed of multiple 6.02-s exposures with associated 0.52-s readout times \citep{gillilandetal2010b}. The LC data integrate over 270 exposures to give 29.4-min observations, whereas the SC data contain nine exposures giving one data point every 58.9\,s. Both cadences are stored on-board and downlinked to Earth roughly every 32\,d, introducing gaps up to $\sim$24\,h in length while the photometer is not collecting data. \textit{Kepler} completes one quarter of its orbit after three downlinks, and must then perform a quarterly roll to keep its solar panels pointing towards the Sun, and its radiator pointing to deep space. \textit{Kepler} data are therefore organized into quarters and thirds around those rolls and downlinks. LC data quarters are denoted by Q$n$, and SC quarters by Q$n.m$ to notify which third (or `month') of that quarter the data correspond to.

Pre-Q9, \textit{Kepler} data were available in two forms: (i) `raw' flux, of which Simple Aperture Photometry (SAP) flux is the preferred nomenclature, and for which basic calibration is performed distinguishing it from the truly raw flux, and (ii) Pre-search Data Conditioned (PDC) `corrected' flux. The PDC data were created as a step towards facilitating planetary transit searches and should be used only with caution in astrophysical analyses because some stellar variability can be modified in the light curves by the PDC pipeline, pertaining to data releases 11 and earlier. This is discussed in Section\,5. Post-Q9, PDC has been superseded by another pipeline, PDC\,MAP. New quarters of data will contain PDC\,MAP rather than the old PDC fluxes, and older quarters of data are due to be reprocessed with this pipeline and made public by July 2012.

There are many advantages of SC data over LC data, but hardware limitations restrict SC slot allocation to 512 slots at any given time. Here we discuss the following advantages of SC data: increased sampling rate; higher Nyquist frequency; fewer low-frequency artefacts; and reduced errors on frequency, amplitude and phase determinations in the Fourier spectrum. We also discuss the difference in distribution of data points between SC and LC data and look at the differences in the PDC and SAP-flux data. Initial characteristics of the LC and SC data can be found in \citet{jenkinsetal2010} and \citet{gillilandetal2010b}, respectively. For a detailed, recent review of \textit{Kepler} noise properties, see \citet{gillilandetal2011} and references therein.

\section{Sampling rate}

There are 30 times more SC points than LC points in a quarter, arising from the longer integration time used for LC data. Two well-known effects this has on time-series analysis are the time resolution available, and the associated Nyquist frequency:

\subsection{Time resolution}

The primary mission goal of planet detection requires the increased sampling rate to time transits more precisely; signs of gravitational perturbation seen as changes in transit duration may lead to subsequent detections of other planets orbiting the same star. \citet{holman&murray2005} calculate that under the gravitational influence of other solar system bodies, Earth's apparent transit time for an observer viewing along the orbital plane would appear to decrease by around 650\,s for 2 in 10 of its orbits, depending on the relative position of each planet and the observer with respect to orbital phase. The effect is even greater for planets orbiting farther from their star (up to $\sim$6000\,s for Mars), and for planets orbiting less massive stars. The sensitivity required for such detections is easily met by \textit{Kepler}, where transit durations would vary by 11 and 101 SC points for the Earth and Mars cases respectively, even where errors from photon statistics on such a transit duration reach $\sim$500\,s, equivalent to 8 SC points \citep{holman&murray2005}. Indeed, the first previously unknown planet to be detected using this technique, Kepler-19c, does not appear to transit its star \citep{ballardetal2011}, but leaves a clear sinusoidal deviation of the transit times once the transits of Kepler-19b are subtracted out.

\begin{figure}
\begin{center}
\includegraphics[width=0.5\textwidth]{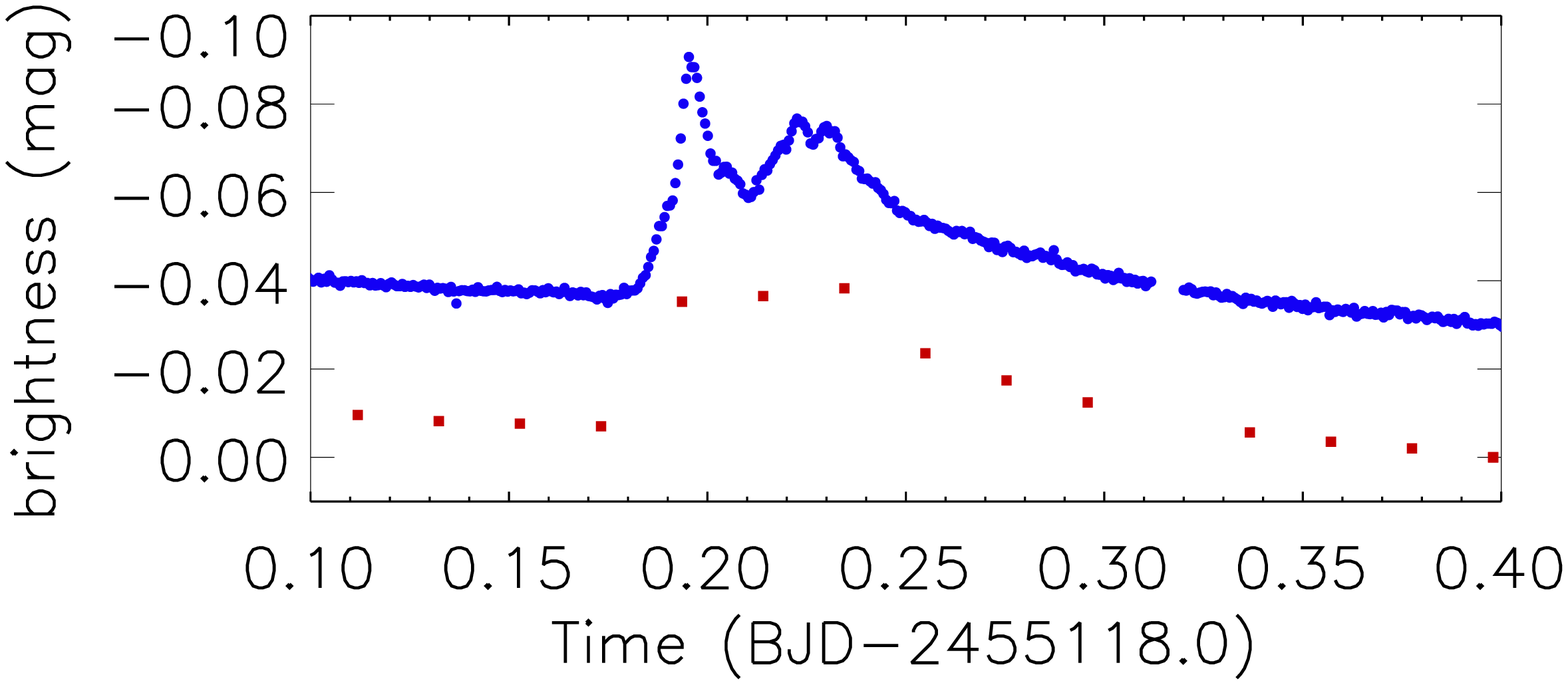}
\caption{A large-amplitude flare on KIC\,12406908. The LC data (red squares) are plotted beneath the SC data (blue circles) for comparison. The change in brightness is precise, but not accurate -- the dimmest LC point was chosen as the zero point for the graph, and all SC points are offset for clarity. The SC data used are Q3.1 PDC flux (see Section\,5 for more details on PDC flux), and the LC data are simulated by averaging bins of 30 consecutive SC points.}
\label{fig:flare}
\end{center}
\end{figure}

The higher sampling rate is useful for astrophysical studies too, particularly for short-lived events like flares. The star KIC\,12406908 is in Kepler Asteroseismic Science Consortium\footnote{http://astro.phys.au.dk/KASC/} Working Group 7 (KASC WG7: Cepheid Variables), but is probably misclassified. It is one of the $\sim$15\,per\,cent of the Kepler Input Catalogue (KIC) stars that have no fundamental parameters listed, that is, no $T_{\rm eff}$ or $\log g$ values are available. Fig.\,\ref{fig:flare} shows a 7.2-h sample of the light curve of the largest flare in Q3.1. The LC data have been plotted underneath the SC data for comparison, and a shift in magnitude has been created for demonstrative purposes only. The longer integration time in LC has the effect of averaging the SC points, and under-samples the 0.056\,mag flare. The shape of the flare, including its erratic nature as its luminosity output rises and falls numerous times across the event, is lost in the LC data. Only a rough approximation of its magnitude and duration would be determinable without the SC data. Events with such short time-scales can clearly only be studied in SC.

\subsection{Nyquist frequency}

The most important benefit of SC data to asteroseismology is the higher Nyquist frequency, and the exquisite quality of the \textit{Kepler} data provide a nice opportunity to demonstrate this. Many asteroseismic targets pulsate at frequencies higher than the Nyquist frequency of the LC data (24.469\,d$^{-1}$; 283.21\,$\mu$Hz), and cannot be studied reliably due to aliasing problems. Solar-like oscillations and roAp star pulsations only occur at frequencies much higher than this. Straddling both sides of the LC Nyquist frequency are the $\delta$\,Sct stars. These stars pulsate in low-order p-modes (pressure modes) and typically have frequencies in the range 4-50\,d$^{-1}$ (46-579\,$\mu$Hz; \citealt{breger00}), although the highest published frequency for a $\delta$\,Sct star is currently 79.5\,d$^{-1}$ (920\,$\mu$Hz; \citealt{amadoetal2004}).

The Nyquist frequency is equal to half the rate at which a signal is being sampled. Since LC has a point every 29.4\,min, there are 48.9 points per day, and the Nyquist frequency is therefore 24.5\,d$^{-1}$ (284\,$\mu$Hz). Hence if in the Fourier spectrum a signal is detected with a frequency higher than the Nyquist frequency, it is not fully sampled, and an alias will be detected at 2$f_{\rm Nyquist} - f_{\rm signal}$. It is not always obvious that these detected frequencies are `reflections' of frequencies higher than $f_{\rm Nyquist}$, and can sometimes be interpreted as real pulsation frequencies.

One is naturally cautious when any frequencies are detected in LC near the Nyquist frequency, as the star could have pulsation frequencies above the Nyquist frequency even if the detected frequencies are real and not reflections. However, for much lower frequencies the possibility of a reflection seems more remote.

\begin{figure*}
\begin{center}
\includegraphics[width=0.9\textwidth]{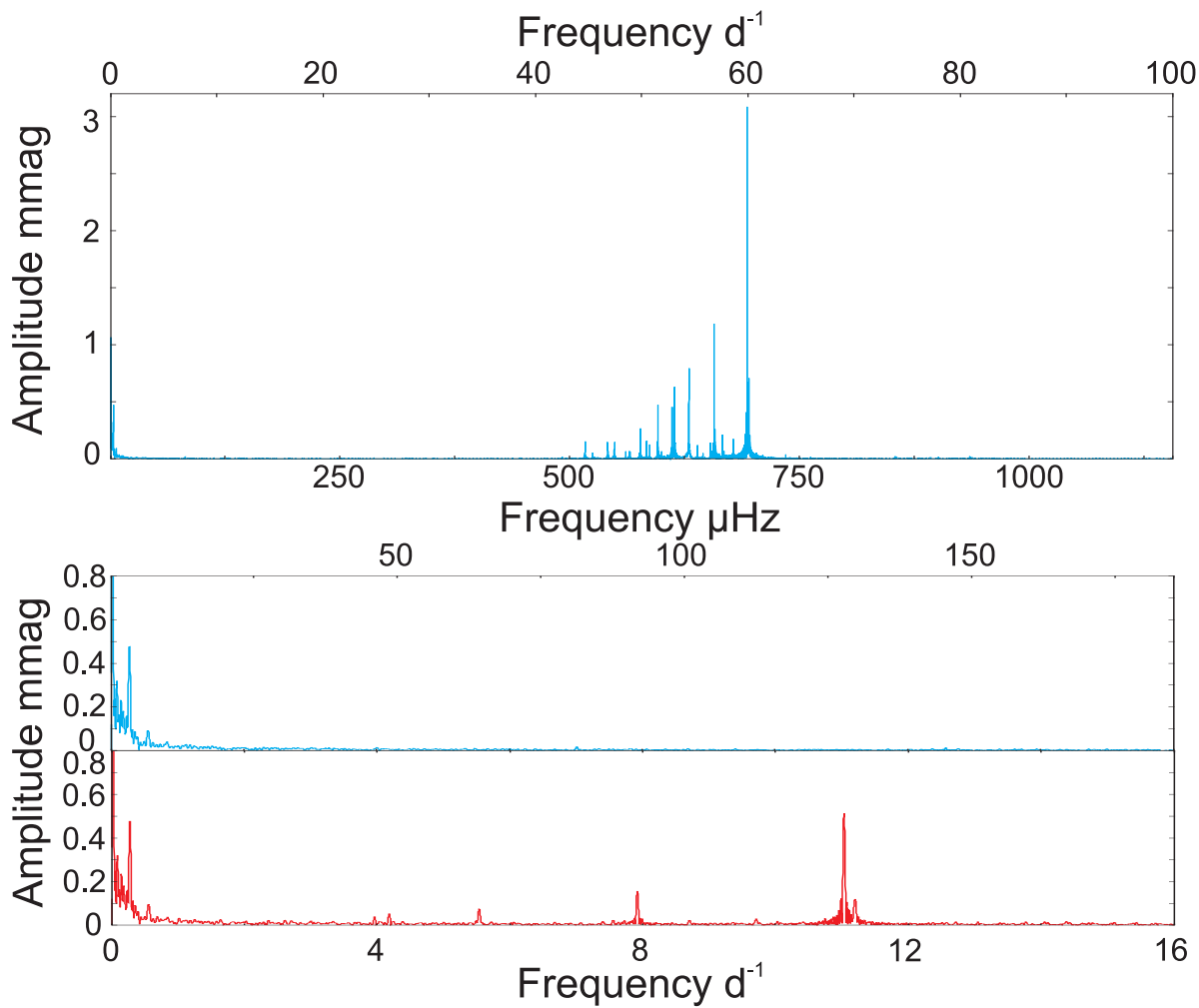}
\caption{\textit{Upper panel}: the Fourier spectrum of the $\delta$\,Sct star KIC\,10977859 for Q1, showing only high-order, high-frequency p-modes, and a lack of pulsations at lower frequencies. \textit{Lower panels}: A magnification of the spectrum between 0 and 16\,d$^{-1}$ for the same star, in SC (blue, middle) and LC (red, bottom). The amplitude of the real peaks in the top panel are up to five times greater than corresponding peaks in the bottom panel, indicating a significant amplitude reduction in reflected peaks. No points were removed from these SAP-flux data, which were analysed using the statistical package Period04 \citep{lenz&breger2004}.}
\label{fig:reflection0-100}
\end{center}
\end{figure*}

KIC\,10977859 is a $\delta$\,Sct star in which the SC data show only high-order p-modes, and a lack of pulsations at frequencies below the LC Nyquist frequency (Fig.\,\ref{fig:reflection0-100}, upper panel). When only the LC data are considered (if SC data were not available, for instance), the spectrum looks entirely different. Nyquist frequency limitations mean that in LC data the true frequencies of high-order p-modes would not be discernible, but a huge number of peaks are visible below the LC Nyquist frequency instead (Fig.\,\ref{fig:reflection0-100}, lower panel), strongly implying the star pulsates in low-order p-modes and maybe g-modes (gravity modes), too. What is even more misleading in this case, and makes the situation problematic, is the absence of peaks in the periodogram of the LC data between 16 and 24.4\,d$^{-1}$ (185-282\,$\mu$Hz), fooling the observer into believing these are low-order p-modes with only a small likelihood of signals at higher frequencies -- there is no warning of what lies beyond the LC Nyquist frequency.

One must exercise extreme caution when analysing LC data if there are no SC data to test for aliasing problems associated with the LC Nyquist frequency.

\section{Low frequency peaks}
Another benefit of SC data over LC is the reduced number of high-amplitude peaks at low frequency. Such peaks can arise naturally from long time-scale processes such as differential velocity aberration, with stars moving across the CCD by up to 1.5 pixels, which results in a different flux fraction being captured by the CCD, and the amount of background contaminating light changing. \citet{garciaetal2011} cite CCD degradation as a cause of long time-scale drifts too, but CCD degradation is likely to arise from high-energy cosmic ray impacts and will often, as a result, be more of a step-function than a long-term trend. SAP-flux data do contain strong instrumental trends that dominate at low frequency, rendering the difference in prevalence of low-frequency peaks between the two cadences insignificant in SAP-flux. However, the LC PDC flux data, from which instrumental trends have mostly been removed, still contain some relatively high-amplitude peaks at low frequency.

\begin{figure*}
\begin{center}
\includegraphics[width=0.9\textwidth]{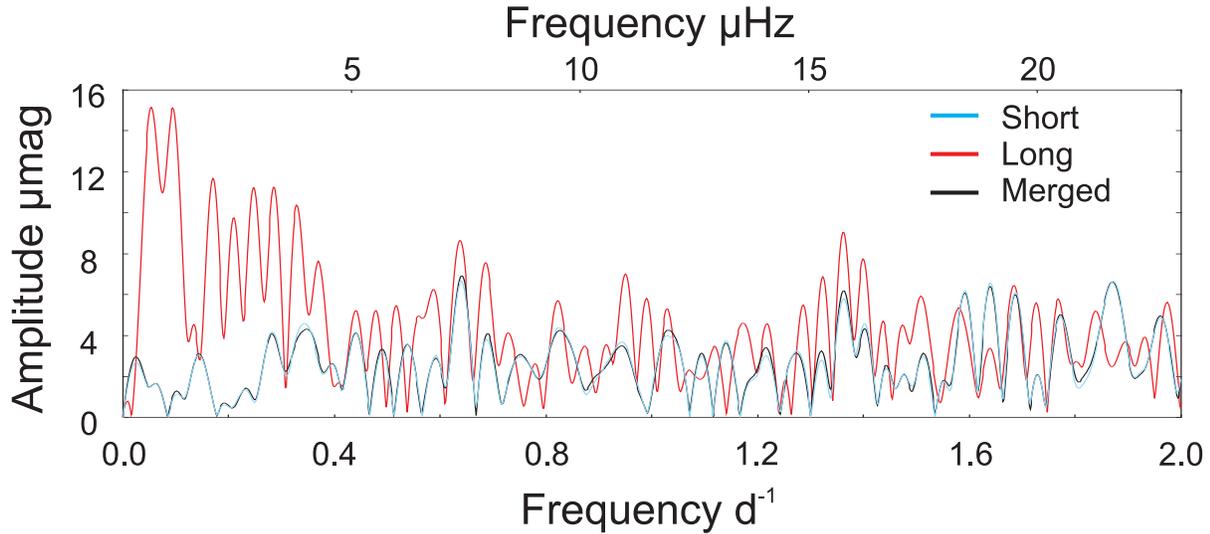}
\caption{The PDC SC data (blue) are plotted on top of the PDC LC data (red) of Q2.2 for the near-constant star KIC\,9390100 for frequencies 0-2\,d$^{-1}$. Plotted in black are the LC data created from merging the PDC SC data. The blue and black lines almost completely overlap, but the displacement of the red line indicates the extent of artificial low-frequency peaks introduced by the PDC data pipeline.}
\label{fig:low_frequency}
\end{center}
\end{figure*}

In order to demonstrate this phenomenon, a nearly constant star was selected to minimize the number of peaks seen as a result of pulsations. The nature of \textit{any} frequency peaks in the Fourier transform depends greatly on the length of the dataset. For one of the stars analysed, KIC\,9390100, the LC Q2 data span 88.9\,d, but the SC Q2.2 data only span 30.0\,d --  the star was not observed in SC during Q2.1 and Q2.3. It was therefore necessary to truncate the length of the LC data to that of the SC Q2.2 data. Regardless of cadence, each quarter is divided into thirds by \textit{Kepler}'s downlinking process, thereby the cadences corresponding to Q2.1 and Q2.3 in the LC data were easily removed. No outliers were removed from either dataset. A third dataset was made for comparison by binning sets of 30 consecutive SC points and replacing them with a single point; the fluxes of the SC points were added and their times averaged. This `merging' process creates points that are entirely concurrent with the LC data (for every LC point there is a merged point at exactly the same time), and the merged and \textit{Kepler} LC datasets shown in Fig.\,\ref{fig:low_frequency} contain exactly the same number of points. To achieve this, two more points had to be removed from the LC dataset, at BJD = 2455042.3243 and 2455051.4377, because undefined SC flux values in these bins did not allow accurate merged data points to be created, but \textit{Kepler} LC points existed at those times. The 29$\times$2 SC data points corresponding to each of these LC points were also removed for a fair comparison in this PDC-data example.

At lower frequencies (below 2\,d$^{-1}$, and especially below 0.4\,d$^{-1}$) the SC and LC spectra are significantly different (Fig.\,\ref{fig:low_frequency}). There are a few coherent peaks common to both cadences, but the LC data have many more high-amplitude\footnote{relatively speaking: amplitudes of 10$^{-5}$\,mag are normally considered tiny in the analyses of $\delta$\,Sct stars!} peaks at those frequencies under 2\,d$^{-1}$. This difference between SC and LC \textit{Kepler} data is often encountered, and requires treating before frequency analysis.

At higher frequencies (above 2\,d$^{-1}$) the two amplitude spectra are almost the same, and can be seen converging in the figure. The merged data, which one expects to be similar to the LC data because they are created by integrating fluxes over the same cadence-numbers, mimic the SC data very well over the entire frequency range. Slight differences can be accounted for by considering that there are occasionally SC data points next to data gaps that do not get incorporated into merged data, because the merged data must be concurrent with the LC data for a direct comparison.

The \textit{Kepler} LC data should not behave exactly like the merged data, because LC and SC data go through a different calibration process (involving such methods as dark/bias subtraction and flat-field removal), but the scale of the difference suggests that discrepancies do remain after the PDC-correction procedure. These discrepancies are not seen to the same extent in the SAP-flux data; the difference in peak heights in the PDC example in Fig.\,\ref{fig:low_frequency} is 12\,$\mu$mag (400\,per\,cent) compared with 6\,$\mu$mag in SAP-flux (2\,per\,cent), noting that instrumental trends dominate in SAP-flux. That some discrepancies remain in the PDC data is not surprising, as they were designed to facilitate planet-finding, not asteroseismology. While the residual peaks contribute significantly to the noise on time-scales important to asteroseismology, planetary transit searches are not greatly affected. \citet{gillilandetal2011} compared noise on 6.5-h time-scales, chosen to be representative of planet transit durations; this corresponds to a frequency of 3.7\,d$^{-1}$, but all three lines in Fig.\,\ref{fig:low_frequency} are already converging at 2\,d$^{-1}$. In fact, a planet orbiting a sun-like star and having a transit time of 1\,d -- a time-scale where the discrepancies are an important source of excess noise -- would have a semi-major axis of 3.4\,AU and a period of 6.3\,y. The detection and confirmation of such a planet is much beyond the design capabilities of \textit{Kepler} unless the mission competes successfully for an extension.

In addition to contributions to the total noise level, the low-frequency artefacts can cause other problems, an example of which is in automatic frequency extraction procedures that select the highest amplitude peak or peaks in the Fourier spectrum to classify a star. If the spectrum is dominated by non-astrophysical low frequency peaks, then false classification may occur and lead to incorrect statistics on both dominant frequencies and amplitudes.

A solution to the problem is on the way. PDC\,MAP sees a $\sim$10-20\,per\,cent improvement in signal-to-noise on the 6.5-h time-scales reported on in \citet{gillilandetal2011} (Jon Jenkins, priv. comm.), and supports their conclusions that the major contributor to the observed excess noise still originates in the stars themselves.

\section{Peak widths, amplitudes and errors}

\begin{figure}
\begin{center}
\includegraphics[width=0.49\textwidth]{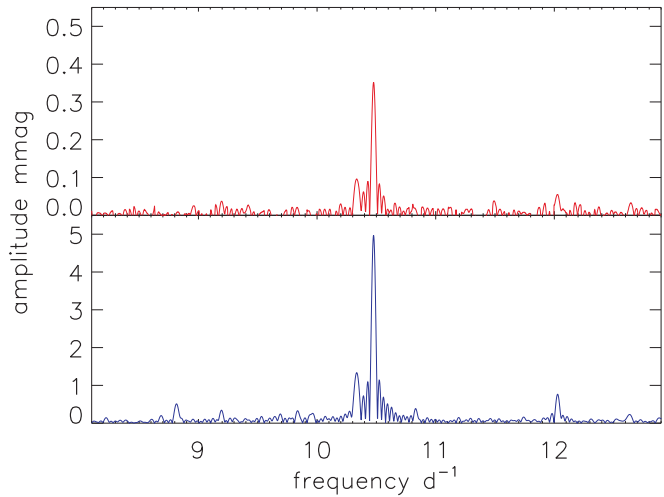}
\caption{The Fourier transform of the LC Q3.2 (truncated) data for KIC\,3437940 is plotted at the bottom in blue, and the \textit{difference} between the amplitudes of the SC and LC data for the same month is plotted above in red, showing that the SC data have slightly higher-amplitude peaks due to the shorter integration time and the averaging effect inherent within LC data. Notice the change in scale on the vertical axis by a factor of 10 between the two plots.}
\label{fig:amplitude_diff}
\end{center}
\end{figure}

The width of a peak in the Fourier spectrum of a dataset of length $T$ can be approximated by $1/T$ (the Rayleigh criterion). For datasets of equal length, the greater number of points in SC data has no effect on the width of the peak, but the amplitude of peaks is different. As mentioned in Section\,2, the longer integration time of the LC data causes an averaging effect that is more noticeable for shorter-period events or pulsations. The same effect reduces the amplitude of peaks in the periodogram of LC data, as shown in Fig.\,\ref{fig:amplitude_diff}. The percentage difference in amplitude between SC and LC increases with higher frequencies.

This effect can be explained mathematically. For a Fourier peak with true amplitude $A_0$, it can be shown (see online supplementary material) that the observed amplitude, $A$, is described by the equation
\begin{equation}
A = \frac{\sin \pi/n}{\pi/n}A_0
\label{eq:amp_reduction}
\end{equation}
for $n$ points per cycle. Using KIC\,3437940 (from Fig.\,\ref{fig:amplitude_diff}) as a numerical example, the ratio of the observed amplitude in SC to that of LC, $A_{\rm SC}/A_{\rm LC}$, is 1.08 for the peak at 10.5\,d$^{-1}$, but $A_{\rm SC}/A_{\rm LC}$ increases to 1.34 for a hypothetical peak at 20\,d$^{-1}$.

KIC\,10977859 (Fig.\,\ref{fig:reflection0-100}) demonstrates the same effect for high-frequency pulsations that are reflected in the Nyquist frequency. The highest peak in the bottom panel is located in frequency just where one expects from the reflection of the highest peak in the top panel (to well within one FWHM), and has an amplitude reduction in agreement with equation (\ref{eq:amp_reduction}) to within the least-squares errors. One can therefore do asteroseismology on p-modes above the LC Nyquist frequency using LC data, providing at least one month of SC data is available to overcome the aliasing problem.

Having a greater number of points allows a more precise determination of pulsation frequencies, amplitudes and phases. To give a quantitative example, one month of LC data was compared to SC data for the star KIC\,3437940 and points were only removed to truncate the LC Q3 dataset to exactly the same time-span as the SC Q3.2 dataset -- as was done in Section\,3 for KIC\,9390100. The errors on frequency, amplitude and phase for the different cadences are summarised in Table\,1, and are on the order of 5 times greater for the LC data. The result applies to both SAP and PDC flux, indicating the precision difference is not a result of the greater variance at low frequency presented in Fig.\,\ref{fig:low_frequency}. This is an important distinction pointing to greater quality of the SC data, and implies the data are not normally distributed. \citet{degrooteetal2009} found a similar result for CoRoT noise properties.

\begin{table}
\caption{The formal least-squares errors on frequency, amplitude, and phase for Q3.2 SC and LC data of the star KIC\,3437940, with the LC errors being greater in each case by factors of $\sim$5. The least-squares errors were calculated with Period04.}
\begin{tabular}{l c c c}
\hline
cadence 	& frequency error	& amplitude error	& phase error\\
		& $\times10^{-5}$ d$^{-1}$ & $\mu$mag	&  $\times10^{-3}$ rad\\
\hline
SC PDC & $ 4.9 $ & $ 14 $ & $ 0.4 $	\\
\vspace{1.5mm}
LC PDC & $ 25.8 $ & $ 70 $ & $ 2.2 $\\
SC SAP & $ 4.6 $ & $ 13 $ & $ 0.4 $	\\
LC SAP & $ 25.8 $ & $ 70 $ & $ 2.2 $\\
\hline
\end{tabular}
\end{table}

The scatter of points in SC data is greater than that of LC data. There are 30 times more points in SC data, so the scatter of points is expected to be $\sqrt{30}$ times greater if the noise is assumed to be white\footnote{The noise is not white, but this serves as a useful approximation.}. This is particularly noticeable in those stars that are approximately constant. In those that pulsate, one has to be careful when discarding those points that appear to be outliers -- there should be more outliers in SC because of the greater number of points, but sampling the brightness variations more often produces higher-amplitude peaks in both the light curve and the Fourier transform. If one `sigma clips' the data too closely from the beginning, the extrema of those peaks in the light curve might be lost. Clipping at 3\,$\sigma$ is too tight -- \textit{if} the data were normally distributed, one discards 1 in 200 points that naturally belong to the distribution in this manner; these points may lie further from the mean or fit, but are not necessarily erroneous outliers. A discussion of the validity of sigma clipping and other outlier removal procedures can be found in \citet{hoggetal2010}.

\section{PDC vs SAP-flux}

Data-files downloaded through either the Kepler Asteroseismic Science Operations Centre (KASOC\footnote{http://kasoc.phys.au.dk/}) or the NASA multimission archive\footnote{http://archive.stsci.edu/kepler/} contain times of observations, SAP-flux, PDC flux (or PDC\,MAP, depending on data release), and the errors on those fluxes. The SAP-flux data show instrumental trends, but the PDC flux light curves correct some of these. In both cases, bad cadences are flagged in the .fits data-file format. \citet{garciaetal2011} discuss the process of correcting \textit{Kepler} light curves in more detail and specifically its application to asteroseismic analysis, even creating their own, separate, pipeline. A more thorough discussion of the PDC pipeline with more general applications can be found in the Kepler Data Characteristics Handbook\footnote{http://archive.stsci.edu/kepler/manuals/
Data\_Characteristics\_Handbook\_20110201.pdf} \citep{christiansen&vancleve2011}. Here, we focus on a few common issues, not including the information concerning processing of LC data that was presented in Section\,3.

\subsection{Instrumental effects}

The most significant effects on \textit{Kepler} light curves are differential velocity aberration (discussed in Section\,3); loss of fine pointing, which leads to small gaps in the light curves in both SAP and PDC flux; reaction wheel zero-crossings, during which the spacecraft shakes for a day or so (not corrected in PDC data); cosmic ray events, which cause a step-function change in the flux level, which decays exponentially back to 90-100\,per\,cent of the original level (corrected in PDC data); monthly Earth downlinks, identifiable by a gap in the data of up to $\sim$24\,h followed by an exponential increase/decrease in flux level as the telescope returns to science operating focus and temperature (corrected in PDC data); and attitude tweaks, which have thus far occurred only twice in the entire mission, during science operations in Q2, and are no longer expected to be a problem (not fully corrected in PDC data). Further details on these effects can be found in the Kepler Data Characteristics Handbook, which also keeps track of lists of known spurious peaks belonging to the SC and LC data.

\subsection{Optimal aperture corrections}
In SAP and PDC flux data alike, the amount of flux in the defined aperture changes from quarter to quarter, and discontinuities in flux are seen as a result. The PDC data contain corrections for both: a) the amount of light contributed from the intended target and not contaminating stars: the median flux over a month or quarter is multiplied by ($1 - \rm{contamination}$) and subtracted from each cadence; and b) the fraction of flux not captured by the `optimal aperture', which is defined to maximise signal-to-noise rather than to capture all light from the star. The SAP-flux data contain no such corrections.

\subsection{Effect on data analysis}
\begin{figure*}
\begin{center}
\includegraphics[width=0.9\textwidth]{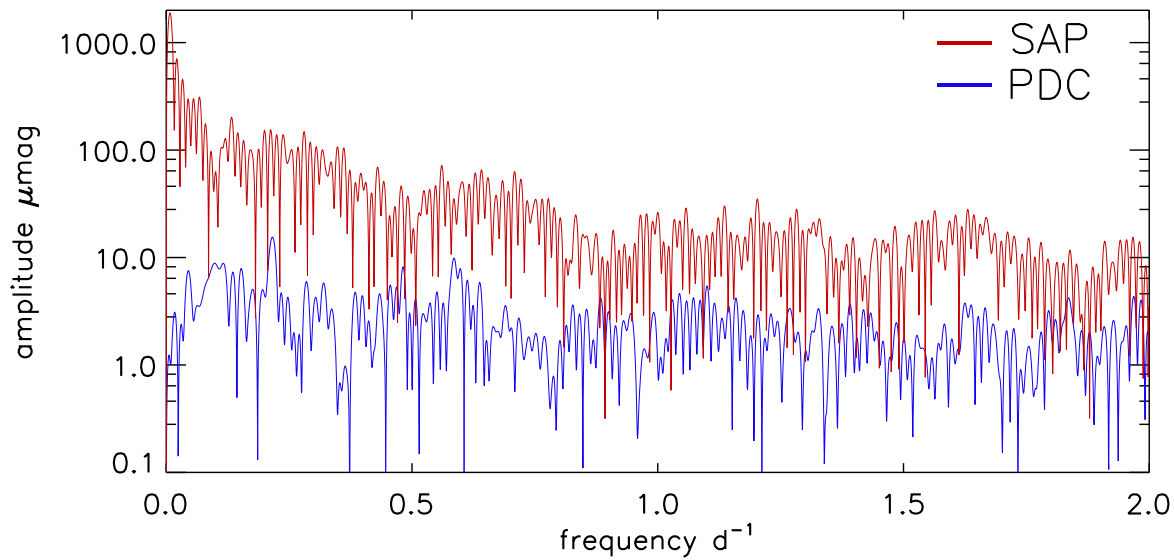}
\caption{The PDC-flux data (blue) have been plotted over the SAP-flux data (red) for the \textit{Kepler} object KIC\,7450391 during Q2. No points were removed from either dataset. The low-frequency peaks in PDC LC data mentioned in Section\,3 are present but insignificant by comparison to the difference between PDC and SAP data.}
\label{fig:corrected_flux}
\end{center}
\end{figure*}

Whilst analysing the effect of cadence on the prevalence of low-frequency peaks in the periodogram, the PDC flux data were compared with the SAP-flux data. Fig.\,\ref{fig:corrected_flux} illustrates the pervasiveness of low-frequency peaks in the SAP data compared with the much flatter PDC flux, for the near-constant star KIC\,7450391, using the full Q2 LC dataset and not removing outliers. Not only are the SAP-flux data visibly poorer in this regard, but the amplitude of the PDC flux curve needed artificial amplification by a factor of ten, simply to make it visible on these axes. This puts some perspective on the problematic LC PDC flux data, in that the SAP-flux data are dominated much more substantially by artificial peaks than are the PDC flux data.

\section{Conclusion}

The SC data are almost always better than the LC data. The SC slot availability is the major limitation one faces when using and obtaining SC data. We have seen the necessity of the increased sampling rate of SC data for resolving short time-scale events such as flares, and for precise transit timing.

The Nyquist frequency of LC data can be problematic and certainly limiting in asteroseismic analysis. Pulsations detected near the Nyquist frequency can be indicative of higher-frequency pulsations that require SC data, and also that peaks in the periodogram may be reflections of peaks from above the Nyquist frequency. However, the lack of pulsations near the Nyquist frequency cannot rule out the presence of higher-order p-modes, and SC data may still be required. When SC data are not available to check for frequencies above the LC Nyquist frequency, one must be aware that observed signals could be reflections from beyond the LC Nyquist frequency.

There is an amplitude difference between SC and LC data associated with the longer integration time of LC data. The result is that peaks in the periodogram have slightly higher amplitudes in SC data, and the percentage difference between the two cadences grows with increasing frequency.

Effects that are not directly concerned with the different sampling rates have also been observed. LC PDC flux data often contain spurious peaks of non-astrophysical origin at very low frequency that are not always present with similar amplitudes in SC PDC data. This can affect studies of long-period brightness variations arising from such things as spots on slow rotators. Automatic frequency extraction can suffer from these artificial peaks.

The greater number of points in SC data does not produce narrower peaks in the periodogram and cannot improve resolution of two closely-spaced frequencies. Frequencies can be determined with greater precision though, as can their corresponding amplitudes and phases. For this to be true, the data cannot be normally distributed.

The usefulness of PDC flux data was also discussed. Noise in the periodogram can be vastly reduced by analysing the PDC flux data instead of SAP-flux data. The PDC flux data have fewer drifts, jumps and outliers, generating cleaner light curves and Fourier spectra, but may also modify astrophysical signals. It is therefore recommended to cross-examine results obtained with PDC flux data with those from SAP-flux data, as advised in the Prefatory Admonition of the Data Characteristics Handbook, to check that genuine pulsation frequencies have not been missed as a result of accidental removal in the data processing pipeline. The PDC flux data do not necessarily remove all jumps and outliers, so it is still recommended to manually check light curves for such artefacts prior to analysis. Moreover, it is recommended that investigators analyse the subset of data points that have a quality flag of zero, meaning the cadences are `good'. Of the observations made during \textit{Kepler}'s 92\,per\,cent duty cycle, $\sim$95\,per\,cent of data points in `well-behaved' quarters might be described as `good' for a typical 13$^{\rm th}$ magnitude star. Data generated in this method will supersede the present PDC flux data for asteroseismic analysis.

Finally, the PDC pipeline leaves artificial peaks in the LC data. There are relatively high-amplitude peaks at low frequency ($<2$\,d$^{-1}$) in the Fourier transform of the LC PDC flux data. Such time-scales are important to asteroseismology, but unimportant for the transits of potentially habitable planets. Never-the-less, improvement is expected in data processed in Quarter 9 Release 12 and subsequent releases, through the PDC\,MAP pipeline. 

\vspace{10 mm}

I would like to thank Don Kurtz for advice and discussions and acknowledge the financial support of the STFC via the PhD studentship programme.

\bibliography{SC_characteristics_arXiv}
\end{document}